\documentclass[12pt]{article}
\usepackage[psamsfonts]{amssymb}
\usepackage{amsmath, amsthm, anysize, enumerate, color, url}
\usepackage{graphicx}
\usepackage{epstopdf}
\usepackage{epsfig}
\usepackage{subfigure}
\usepackage[ruled,linesnumbered]{algorithm2e}

\usepackage[sectionbib]{natbib}

\usepackage{threeparttable}

\usepackage[colorlinks, breaklinks,
                   linkcolor=red,
                  citecolor=blue
                  ]{hyperref}

\usepackage{breakurl}
\usepackage{xurl}

\usepackage{bm}
\usepackage{booktabs}

\newtheorem{theorem}{Theorem}
 
\newtheorem{corollary}{Corollary}

\newtheorem{lemma}{Lemma}

\theoremstyle{definition}
\newtheorem{condition}{Condition}

\newcommand{\R}{\mathbb{R}}

\renewcommand{\P}{\mathbb{P}}

\newcommand{\diag}{\mbox{diag}}
\newcommand{\bs}{\boldsymbol}

\usepackage{setspace}

\begin{document}

\title{Semiparametric analysis for paired comparisons
 with covariates}

\author{Haoyue Song\thanks{Department of Statistics, Central China Normal University, Wuhan, 430079, China.
\texttt{Email:} hysong05@mails.ccnu.edu.cn.}
\hspace{4mm}
Lianqiang Qu\thanks{Department of Statistics, Central China Normal University, Wuhan, 430079, China.
\texttt{Email:} qulianq@ccnu.edu.cn.}
\hspace{4mm}
Ting Yan\thanks{Department of Statistics, Central China Normal University, Wuhan, 430079, China.
\texttt{Email:} tingyanty@mail.ccnu.edu.cn.}
\hspace{4mm}
Yuguo Chen\thanks{Department of Statistics, University of Illinois Urbana-Champaign, Champaign, IL 61820, USA.
\texttt{Email:} yuguo@illinois.edu.}
\hspace{4mm}
}
\date{}

\maketitle

\begin{abstract}
\begin{spacing}{1.2}
Statistical inference in parametric models (e.g., the Bradley--Terry model and its variants)
for paired-comparison data has been explored in the high-dimensional regime,
in which the number of items involving in paired comparisons diverges.
However, parametric models are highly susceptible to model misspecification.
To relax the assumption of known distributions and provide flexibility,
we propose a semiparametric framework for modeling the merits of items and covariate effects (e.g., home-field advantage)
by introducing latent random variables with unspecified distributions.
As the number of parameters increases with the number of items, semiparametric inference is highly nontrivial.
To address this issue, we employ a kernel-based
least squares approach to estimate all unknown parameters.
When each pair of items has a fixed number of comparisons and the number of items tends to infinity,
we prove the consistency of all resulting estimators and derive their asymptotic normal distributions.
To the best of our knowledge, this is the first study to conduct a semiparametric analysis of paired comparisons with an increasing dimension.
We conduct simulations to evaluate the finite-sample performance of the
proposed method and illustrate its practical utility by analyzing an NBA dataset.
\end{spacing}

\vskip 5 pt \noindent
\begin{spacing}{1.4}
\end{spacing}

\end{abstract}

\vskip 5pt

\section{Introduction}

Paired comparison is a popular approach for ranking a set of items,
where the preference between two items is judged each time. This technique is especially useful when the evaluation criteria are objective by nature. Paired-comparison data are naturally generated in sports tournaments, where multiple teams compete against each other. A central problem in paired-comparison analysis is quantitatively characterizing
the merits of items to derive a ranking. In balanced-comparison designs, where each pair of items has the same number of comparisons, the ranking is based on the win counts. However, in unbalanced-comparison designs, there is no natural ranking. To address this issue, statistical models, such as the Bradley--Terry \citep{bradley-terry1952} and Thurstone \citep{thurstone1994a} models,
are employed to estimate the merit parameters of items in unbalanced comparisons, and rank is produced according to their estimated values. The applications of paired-comparison include the ranking of classical sports teams \citep{masarotto2012the, Sire_2008baseball, Whelan-2020-Hockey}, scientific journals \citep{stigler1994citation, Varin-2016-jrsa}, product brands \citep{radlinski2007active}, and crowd-sourced labels \citep{chen2016overcoming}.

Classical paired-comparison models, including the Bradley--Terry and the Thurstone models, assign
a merit parameter $\theta_i$ to each item and assume that the probability of one item defeating another item depends only on the relative difference in their merit parameters.
As argued by \citeauthor{David1988} (\citeyear{David1988}, page 7), the merit of an item $i$ can be represented by a latent random variable $\epsilon_i$,
capturing uncertainty across observations.
In a paired comparison of items $i$ and $j$, items $i$  wins if $\theta_i - \theta_j > \epsilon_i-\epsilon_j$;
otherwise, items $j$  wins.
Parametric models require the distribution of $\epsilon_i$ to be known.
When $\epsilon_i$ follows a normal or doubly exponential distribution, it corresponds to the Thurstone or Bradley--Terry model
\citep[page 8]{David1988}, respectively.
However, the distribution of $\epsilon_i$ is usually unknown. Hence, it may lead to incorrect inference if not correctly specified.
Moreover, covariate information may accompany paired-comparison data.
For example, a team is more likely to win when playing in its home city in most sports,
which is referred to as the home-field advantage \citep[e.g.][page 455]{agresti2012}. 
The winning rate of a team in the previous regular season could also be used to predict its performance in the current season.
Ignoring the covariate effects may also lead to invalid inference, as demonstrated in \cite{yan2020paired}.

In this study, we propose a general semiparametric framework for modeling the merits of items and covariate effects, when the distribution of the latent random variable is unknown.
Let $\bm{X}_{ijt}=(X_{ijt,0},X_{ijt,1},\ldots,X_{ijt,p})^{\top}$  denote the covariate information
associated with the $t$th comparison  of item $i$ against $j$, where $p$ is fixed.
We require $\bm{X}_{ijt}=-\bm{X}_{jit}$ because
if something is advantageous to $i$, then it is disadvantageous to $j$.
We adopt a random-effects design for the covariates $\bm{X}_{ijt}$, that is, $\bm{X}_{ijt}$ are random vectors. Let $a_{ijt}$  indicate whether item $i$ defeats item $j$ in the $t$th comparison:
$a_{ijt}=1$ if item $i$ defeats item $j$ and 0 otherwise.
The semiparametric framework assumes that the winning probability of $i$ against $j$ conditional on $\bm{X}_{ijt}$ is
\begin{eqnarray}\label{model}
\P( a_{ijt}=1|\bm{X}_{ijt}, \bs{\gamma}, \theta_i, \theta_j ) = F(  \theta_i - \theta_j + \bm{X}_{ijt}^\top \bs{\gamma}) ,
\end{eqnarray}
where $\bs{\gamma}$ is a $(p+1)$-dimensional regression coefficient of the covariates, $\theta_i$ is the merit parameter of item $i$.
Here, $F(x)$ denotes the cumulative distribution function of latent random variables that satifying $F(x)+F(-x)=1$.
Under the restriction $\bm{X}_{ijt}=-\bm{X}_{jit}$, the probability distribution in \eqref{model} is well defined.
Hereafter, we call it the semiparametric paired-comparison model.
When $F(x)$ is the cumulative distribution function of a logistic or standard normal variable, it becomes the Bradley--Terry or Thurstone model, respectively.

The covariate $\bm{X}_{ijt}$ can be defined based on the situations of the teams or the items' attributes.
If $\bm{W}_{it,j}$ and $\bm{W}_{jt,i}$ denote $q$-dimensional attributes of items $i$ and $j$ in the $t$th comparison between items $i$ and $j$, respectively,
they can be used to construct the  vector $\bm{X}_{ijt}=\bm{g}(\bm{W}_{it,j}, \bm{W}_{jt,i})$, where
$\bm{g}(\bm{x}, \bm{y})=-\bm{g}(\bm{y},\bm{x})$.
For instance, if we let $\bm{g}(\bm{W}_{it,j}, \bm{W}_{jt,i})$ equal to $\bm{W}_{it,j} - \bm{W}_{jt,i}$, then it measures the dissimilarity between the two items.
As an example with a one-dimensional covariate, consider the $t$th game where team $i$ plays at home against team $j$ (the away team), then we let $W_{it,j} = 1$ and $W_{jt,i} = 0$, such that $X_{ijt} = 1$ and $X_{jit} = -1$.

We obtain the conditions for model identification employing a special regressor method \citep{lewbel1998semiparametric,lewbel2000semiparametric}.
Motivated by \cite{qu2026inference}, we develop a kernel-based least squares approach for estimating the unknown parameters in semiparametric paired-comparison models.
We employ a projection matrix to project the covariates onto the subspace spanned
by the column vectors of the design matrix made up of merit parameters, and obtain an explicit estimator for the covariate parameter.
We then estimate the merit parameters by using a least squares method.
The projection procedure eliminates the potential bias of the estimator for the covariate parameter.
We establish consistency and asymptotic normality of the resulting estimators, when the number of items approaches infinity and each pair has a fixed number of comparisons.
Numerical studies and a real-world data analysis demonstrate our theoretical findings.

\subsection{Literature review}\label{Sec1.1}

The Bradley--Terry model and its generalized versions have been examined in the high-dimensional regime,
where the number of items is large.
\cite{simons-yao1999} established the consistency and asymptotic normality of the maximum likelihood estimator (MLE)
in the Bradley--Terry model under a dense comparison assumption, in which each pair has a fixed number of comparisons.
\cite{yan2012sparse} and \cite{Han-chen2020}  generalized the result of \cite{simons-yao1999} to a  sparse comparison case where paired comparisons exist only for some pairs.
\cite{chen2019} found that the spectral method or the
regularized MLE is minimax optimal in terms of the sample complexity---that is, the number of paired comparisons needed to ensure exact top-$K$ identification.
\cite{chen2022optimal}
proved that the MLE achieved optimal partial and exact recovery for the top-$K$ ranking problem, while the spectral method is, in general, sub-optimal.
\cite{han2023general} proved consistency of the MLE in a class of generalized Bradley--Terry models.
\cite{yan2025likelihood} established Wilks-type results for likelihood ratio tests under some increasing and fixed dimensional null hypotheses.
\citet{hunter2003mm} developed a minorization-maximization method to obtain the MLE under generalized Bradley--Terry models.

Recently, \cite{fan2024uncertainty} extended the Bradley--Terry model to incorporate the covariate information
and derived the asymptotic properties of the MLE under an Erd\"{o}s--R\'{e}nyi comparison graph,
where the covariate $X_i$ of item $i$ enters the model additively.
However, this model characterizes only the individual level covariate information and does not address those covariates associated
with each paired comparisons (e.g.,  home-field advantage).
\cite{singh2025least} obtained least squares estimators for cardinal paired-comparison data
with  covariates and established their large sample theory.
\cite{fan2025ranking} studied the ranking problem in a modified version of the Plackett-Luce model for the top choice of multiway comparisons.
\cite{dong2024statistical} explored MLE in a covariate-assisted Plackett--Luce model.
However, these works are based on parametric models, while we focus on semiparametric inference for paired-comparison data with covariates,
in which the error distribution is unspecified.

Paired-comparison data can be represented as a weighted directed graph, where nodes denote items,
and a weighted directed edge from the head node $i$ to the tail node $j$
denotes the number of wins by $i$ over $j$ out of all their comparisons.
Different from the concept of merit parameters in paired comparisons, degree heterogeneity parameters are used to measure the intrinsic nodal merits in forming network connections.
Several parametric models have
been proposed to characterize degree heterogeneity and covariate effects in networks
\citep[e.g.,][]{Graham2017, Yan-Jiang-Fienberg-Leng2018, Dzemski2019},
where the estimation and inference methods depend on a specified distributional assumption for the edges.
To relax the known distribution assumption, semiparametric models have also been developed
\citep{toth2017semiparametric, zeleneev2020identification, candelaria2020semiparametric,qu2026inference}.
Our study is motivated by \cite{qu2026inference}, which introduced a semiparametric framework for directed network formation.
However, they focus on in- and out-degree heterogeneity and homophily effects, which are different from paired-comparison data.

The remainder of this paper is organized as follows. Section \ref{Sec2} presents the semiparametric paired-comparison model, establishes identification conditions, and develops a kernel-based least squares estimator. Section \ref{Sec3} establishes the consistency and central limit theorem for the proposed estimator. Section \ref{Sec4} contains numerical studies and a real-data application.  Section \ref{Sec5} gives some  summarizes. The technical details and additional numerical results are in the Supplementary Material.

{\bf Notations:} Let $[n]_0:=\{0,1,\ldots,n\}$ and
$[n]:=\{1,\ldots,n\}$.
Let $\bm{e}_{i}$ be an $n$-dimensional standard basis vector with the $i$th element $1$
and $0$ otherwise for $i = 1,\ldots, n$.
Further, we define $\bm{e}_{0}=(0, \ldots, 0)^\top$ as the $n$-dimensional zero vector.
Define $\bm{1}_n$ as the $n$-dimensional vector with all elements equal to $1$.
For any $\bm{x}=(x_1,\ldots,x_n)^{\top}\in \mathbb{R}^{n}$, we define the $\ell_2$-norm $\|\bm{x}\|_{2}=\sqrt{\sum_{i\in[n]}x_{i}^2}$ and  the $\ell_\infty$-norm $\|\bm{x}\|_{\infty}=\max_{i\in [n]}|x_{i}|$. For any $k \in [p+1]$,  we define $\bm{x}_{(-k)}=(x_1,\ldots,x_{k-1},x_{k+1}, \ldots,x_n)^{\top}\in \mathbb{R}^{n-1}$.
Let $\bm{I}_{n\times n}$ be the $n\times n$ identity matrix with $1$'s on the diagonal and $0$'s elsewhere, and $\mathbb{I}(\cdot)$ denote the indicator function.
For any matrix $\bm{A}$, we define $\lambda_{max}(\bm{A})$ and $\lambda_{min}(\bm{A})$ as its largest and smallest eigenvalues, respectively. For a matrix $\bm{A}=(A_{ij})_{n\times m} \in \mathbb{R}^{n\times m}$, we define
$\|\bm{A}\|_{\max}=\max_{i\in[n],j\in[m]}|A_{ij}|$, $\|\bm{A}\|_{2}=\sqrt{\lambda_{\max}(\bm{A}^\top \bm{A})}$.
For the positive sequences $\{a_n\}$ and $\{b_n\}$, we write $a_n = o(b_n)$
if $a_n/b_n \rightarrow 0$, as $n \rightarrow \infty$; and $a_n = O(b_n)$ if there exists a constant $C$, such that $a_n \leq C b_n$ for all $n$.
We use the superscript ``$*$" to denote the true parameter, under which the data are generated.

\section{Model, Identifiability and Estimation}\label{Sec2}
\renewcommand{\theequation}{2.\arabic{equation}}
\setcounter{equation}{0}

\subsection{Semiparametric paired-comparison model}
\label{subsec-model}

Assume that there are $n+1$ items, labeled as $0, 1, 2, \ldots, n$, involving in the paired comparisons.
Let $T_{ij}$ denote the number of comparisons between items $i$ and $j$.
For convenience, let $T_{ii} = 0$ for $i\in [n]_0$.
For easy exposition, we set $T_{ij}=T$ for all $0\le i\neq j\le n$, where $T$ is a fixed positive integer.
Our results can be easily extended to the case $K_1\le T_{ij}\le K_2$ for two fixed positive integers $K_1$ and $K_2$.
Define $T_i=\sum_{j=0,j\neq i}^n T_{ij}$ as the total number of comparisons of item $i \in [n]_0$.
Let $a_{i}$ be the number of wins for item $i$, defined as the sum of all its comparison outcomes: $a_{i}=\sum_{j=0, j\neq i}^n \sum_{t\in [T]}a_{ijt}$.

Recall that $\bm{X}_{ijt}=(X_{ijt,0},X_{ijt,1},\dots,X_{ijt,p})^\top\in\mathbb{R}^{p+1}$
denotes the covariate information
associated with the $t$th comparison of item $i$ against $j$, which may affect the outcomes.
We assume that  $a_{ijt}$ are
conditionally independent across $0\leq i< j \leq n$ and $1\leq t\leq T$, given all the covariates $\{\bm{X}_{ijt}\}_{i,j,t}$.

According to the semiparametric paired-comparison model specified in \eqref{model}, $a_{ijt}$ can be represented as
\begin{eqnarray}
\label{Model}
a_{ijt}=\mathbb{I}(\theta_i-\theta_j+\bm{X}_{ijt}^{\top} \bm{\gamma} >\varepsilon_{ijt})~~\text{for}~~i< j~~\text{and}~~t\in [T],
\end{eqnarray}
where $\theta_i~(i\in [n]_0)$ are unknown merit parameters,
$\bm{\gamma}=(\gamma_0,\gamma_1,\ldots,\gamma_p)^\top$ is the regression coefficient vector for the covariates,
and $\varepsilon_{ijt}$ denotes the latent noise.
Under model \eqref{Model}, we observe that the larger the parameter $\theta_i$ is,
the more likely for item $i$ to win. Therefore, $\theta_i$ describes the merit of item $i$  in paired comparisons.
The parameter $\bm{\gamma}$ captures the covariate effects on the comparison outcomes.
In the case of home-field advantage, if the parameter $\gamma>0$, item $i$ has a larger probability to win.
The noise term $\varepsilon_{ijt}$ represents unobservable factors that influence the comparison outcomes.
The distribution of $\varepsilon_{ijt}$ is left unspecified.

\subsection{Identifiability and Estimation}

In this section, we first address the identifiability problem of model \eqref{Model},
which is defined over
the joint distribution of covariates and the noises.
Clearly, for any $c_1>0$ and $c_2 \in \mathbb{R}$,
we have
\begin{align*}
a_{ijt}=&\mathbb{I} ({\theta}_i-{\theta}_j + \bm{X}_{ijt}^{\top}{\bm{\gamma}}> {\varepsilon}_{ijt})\\
=&\mathbb{I}\{(c_1\theta_i+c_2)-(c_1\theta_j+c_2) +c_1\bm{X}_{ijt}^{\top}\bm{\gamma}> c_1{\varepsilon}_{ijt}\}.
\end{align*}
Therefore, model \eqref{Model} is unidentifiable without constraints. A common approach to avoid this issue is to set $\sum^{n}_{i=0} \theta_i = 0$
or $\theta_0 = 0$ and $\gamma_k = 1$, where $\gamma_k$ is the $k$th component of $\bm{\gamma}$ and $k$ is chosen such that $X_{ijt,k}$ is a continuous random variable.

The identifiability of model \eqref{Model} is also related to the support of the joint distribution of $(\bm{X}_{ijt},\varepsilon_{ijt})$.
For illustration, we consider a toy example.
Let $\bm{X}_{ijt} \in \mathbb{R}$ be a random variable with  support $(-5, -2) \cup  (2, 5)$.
We set $\bm\gamma = \widetilde{\bm\gamma} = 1$, $\theta_i-\theta_j = 1$ for $0\leq i \neq j \leq n$, and  $\widetilde{\theta_i}-\widetilde{\theta_j}  = c_1(\theta_i-\theta_j) $, where $c_1 \in (0, 1)$. Further, let $\varepsilon_{ij}$ and $\widetilde{\varepsilon}_{ij}$ be obtained from the uniform distribution on $(-1, 1)$. In this scenario, we have
\begin{eqnarray}
\nonumber
&\bm X_{ijt}> \varepsilon_{ij}-1~~~\text{if}~~\bm X_{ijt}\in (2,5)~~\text{and}~~\bm X_{ijt}< \varepsilon_{ij}-1~~~\text{otherwise},
\\ \nonumber
&\bm X_{ijt}> \widetilde{\varepsilon}_{ij}-c_1~~\text{if}~~\bm X_{ijt}\in (2,5)~~\text{and}~~\bm X_{ijt}< \widetilde{\varepsilon}_{ij}-c_1~~\text{otherwise}.
\end{eqnarray}
This implies that $a_{ijt} = \widetilde{a}_{ijt}$,
where $\widetilde{a}_{ijt}=\mathbb{I}\{\widetilde \theta_i-\widetilde\theta_j+\bm{X}_{ijt}^{\top}\widetilde{\bm{\gamma}}> \widetilde\varepsilon_{ij}\}.$
Therefore, model \eqref{Model} cannot be identified in the parameter set $\{0< \theta_i - \theta_j < 1,~ 0 \leq i \neq j \leq n\}$. However, if we change the support of $\bm X_{ijt}$ to $(-5, 5)$, then the support of
$\theta_i - \theta_j - \varepsilon_{ij}$ is a subset of $(-5, 5)$.
Consequently, $\mathbb{P}(a_{ijt} \neq \widetilde{a}_{ijt} ) > 0$
when $\theta_i - \theta_j\neq \widetilde{\theta}_i - \widetilde{\theta}_j$.
In this case, model \eqref{Model} is identifiable.
This phenomenon is also observed in
semiparametric network formation models \citep{qu2026inference}.

We consider the following conditions for the identifiability of model \eqref{Model}.

\begin{condition}
\label{conC}
There exists one continuous covariate $X_{ijt,k}$ such that its regression coefficient $\gamma_k$ is positive.
Additionally, the conditional distribution of $X_{ijt,k}$ given $X_{ijt,(-k)}$ is absolutely continuous with respect to the Lebesgue measure with
nondegenerate conditional density $f(x|X_{ijt,(-k)})$, where
$X_{ijt,(-k)}=(X_{ijt,0},\dots,X_{ijt,k-1}, X_{ijt,k+1},\dots,X_{ijt,p})\in \mathbb{R}^p$.
\end{condition}

The covariate $X_{ijt,k}$ defined in Condition \ref{conC} is referred to as a special regressor \citep{lewbel1998semiparametric, candelaria2020semiparametric, qu2026inference}.
For simplicity, we denote the special regressor as $X_{ijt,0}$ and its conditional density as $f(x|\bm{Z}_{ijt})$, where $\bm{Z}_{ijt}=(X_{ijt,1}, \ldots,X_{ijt,p})^{\top}\in \mathbb{R}^{p}$.
Denote the regression coefficient of the covariate $\bm{Z}_{ijt}$ by
$\bm{\eta}= (\gamma_1,\ldots,\gamma_p)^\top$,
which is identical to $\gamma$ excluding its first element.
For convenience, denote
\begin{equation}
\label{definition-Z}
\overline{\bm{Z}}=\frac{1}{T}\sum_{t\in [T]}\bm{Z}_{t},~~ \bm{Z}_{t} =(\bm{Z}_{01t}, \ldots, \bm{Z}_{0nt},
\bm{Z}_{12t}, \ldots, \bm{Z}_{1nt}, \ldots,  \bm{Z}_{(n-1)nt})^\top \in \mathbb{R}^{N\times p},
\end{equation}
where $N=n(n+1)/2$.

\begin{condition}
\label{conS}
The conditional density $f(x|\bm{Z}_{ijt})$ of $X_{ijt,0}$ given $\bm{Z}_{ijt}$ has support $(-B_U,B_U)$, where $B_U > 0$. Furthermore,
the support of $-(\theta_i^*-\theta_j^*+\bm{Z}_{ijt}^{\top}\bm{\eta}^*-\varepsilon_{ijt})/\gamma^*_0$ is a subset of $(-B_U,B_U)$.
\end{condition}
Condition \ref{conS} restricts the support for $X_{ijt,0}$, which is mild and has been widely adopted \cite[e.g.][]{Man85,lewbel1998semiparametric, lewbel2000semiparametric,candelaria2020semiparametric}. When there are more than one ``special regressor", based on the support requirement in Condition 2, we can use the covariate with the largest observed support among all possible candidates as the special regressor. This is a simple principle that can be easily carried out. Conditions \ref{conC} and \ref{conS} do not impose any restrictions on the distribution of $\bm{Z}_{ijt}$. Therefore, this identification strategy accommodates
discrete covariates in $\bm{Z}_{ijt}$.

\begin{condition}\label{conE}
$\varepsilon_{ijt}$ is independent of $\bm{X}_{ijt}$, and satisfies $\varepsilon_{ijt}=-\varepsilon_{jit}$ and $\mathbb{E}(\varepsilon_{ijt})=0$ $(0\leq i< j \leq n,\,1 \leq t\leq T)$.
\end{condition}
Recall that $\varepsilon_{ijt}$ has a symmetric distribution assumption. Condition \ref{conE} is mild and can be relaxed to the scenario where $\varepsilon_{ijt}$ is conditionally independent
of $X_{ijt,0}$  given $\bm{Z}_{ijt}$.

Next, we introduce some notations. Recall that there are $n+1$ items involved in paired comparisons and we set $\theta_0=0$ for model identifiability.
Let
\begin{eqnarray}\label{def-U}
\bm{U}=(\bm{U}_{01},\ldots,\bm{U}_{0n},\bm{U}_{12},\ldots,\bm{U}_{1n},\ldots,\bm{U}_{(n-1)n})^{\top}\in\mathbb{R}^{N\times n}
\end{eqnarray}
be the design matrix for the parameter vector $\bm{\theta}$, where
$\bm{U}_{ij}=\bm{e}_i-\bm{e}_j\in \mathbb{R}^{n}$
and $\bm{e}_i\in \mathbb{R}^{n}$ is the standard basis vector with the $i$th element $1$ and others $0$. For convenience, define  $\bm{e}_0$ as an $n$-dimensional column vector with all elements $0$.
Denote
\begin{eqnarray}\label{def-V}
\bm{V}=(v_{ij})_{n\times n}:=\bm{U}^{\top}\bm{U}=(n+1) \bm{I}_{n\times n}-\bm{1}_n\bm{1}_n^{\top}\in \mathbb{R}^{n\times n},
\end{eqnarray}
where
$v_{ii}=n, i=1,\ldots,n$ and  $v_{ij}=v_{ji}=-1, 1\leq i\neq j\leq n$.
Here, $\bm{I}_{n\times n}$ denotes an $n\times n$ identity matrix
and $\bm{1}_n$ an $n$-dimensional column vector with all its elements $1$.
Thus, $\bm{V}$ can be viewed as the ``reduced graph Laplacian" obtained by removing the first row and the first column of the graph Laplacian corresponding to the comparison graph of all $n+1$ items, where we assume that each pair is comparisons.
The inverse matrix $\bm{V}^{-1}$ has an explicit expression:
\begin{eqnarray}\label{Vsolve}
\bm{V}^{-1}= \frac{1}{n+1}(\bm{1}_n\bm{1}_n^{\top}+ \bm{I}_{n\times n}).
\end{eqnarray}
Define a projection matrix $\bm{D}$ indicating the orthogonal projection onto the column space of $\bm{U}$:
\begin{equation}
\label{definition-D}
\bm{D}=\bm{I}_{N\times N} -\bm{U}\bm{V}^{-1}\bm{U}^{\top}\in \mathbb{R}^{N\times N},
\end{equation}
where the diagonal elements of $\bm{D}$ are $1-2/(n+1)$, and the off-diagonal elements are either $1/(n+1)$, $-1/(n+1)$, or $0$.
Note that
\[
\bm{DU}=(\bm{I}_{N\times N} -\bm{U}\bm{V}^{-1}\bm{U}^{\top})\bm{U}=\bm{U}-\bm{U}=\bm{0}\in \mathbb{R}^{N\times n}.
\]

\begin{condition}\label{conU}
There is a constant $\lambda>0$ such that $\lambda_{min}(\overline{\bm{Z}}^{\top}\bm{D}\overline{\bm{Z}}/N)\geq \lambda$ almost surely.
\end{condition}

Since $\bm{D}$ has an explicit expression, a direct calculation gives
\[
\overline{\bm{Z}}^{\top}\bm{D}\overline{\bm{Z}}
= \sum_{i<j} \overline{\bm{Z}}_{ij} \overline{\bm{Z}}_{ij}^{\top}
- \frac{1}{n+1} \sum_{i=0}^n \big(\sum_{j,k=0,j,k\neq i}^n \bm{\overline{Z}}_{ij}\bm{\overline{Z}}_{ik}^\top \big),
\]
whose detailed proofs is in 
the Supplementary Material.
A sufficient condition to guarantee Condition \ref{conU}
is that $Z_{ijt}$'s are independently generated from
 a  $p$-dimensional non-degenerated multivariate symmetric distribution.
Let $Z$ denote a general random vector from this distribution.
Under this condition, by the large sample theory,
$N^{-1}\overline{\bm{Z}}^{\top}\bm{D}\overline{\bm{Z}}$
converges to $\mathrm{Cov}(Z)$ almost surely, which implies Condition \ref{conU}.
An intuitive explanation for Condition \ref{conU} is that the amount of information
of the covariate matrix $\overline{\bm{Z}}$ projected into $U^+$ is linearly
proportional to the total amount of information, where $U^+$ denotes the orthogonal complementary space of
the linear subspace $\{\bm{U}\mathbf{x}: \mathbf{x}\in \R^p\}$.
The i.i.d. assumption implies that this condition holds automatically.
In addition, if
 $\overline{\bm{Z}}$ lies in $\{\bm{U}\mathbf{x}: \mathbf{x}\in \R^p\}$,
 then $\bm{D}\overline{\bm{Z}}=0$ such that
 Condition \ref{conU} fails.
Condition \ref{conU} is related to the asymptotic behavior of an estimator of $\bm{\eta}$.
If $\lambda_{min}(\overline{\bm{Z}}^{\top}\bm{D}\overline{\bm{Z}}/N)$ is close to zero, then the asymptotic covariance matrix of $\widehat{\bm{\eta}}$ is ill-posed, which prevents the estimator from retaining good properties.

We next establish the identifiability of the parameters.
Motivated by \cite{qu2026inference},
we define
\begin{eqnarray}\label{Y}
Y_{ijt}=\frac{a_{ijt}-\mathbb{I}(X_{ijt,0}>0)}{f(X_{ijt,0}|\bm{Z}_{ijt})}.
\end{eqnarray}
Further, denote \[\overline{\bm{Y}}=(\overline{Y}_{01},\ldots,\overline{Y}_{0n},\overline{Y}_{12},\ldots,\overline{Y}_{1n},\ldots,\overline{Y}_{(n-1)n})^{\top} ,~~  \overline{Y}_{ij}
=\frac{1}{T}\sum_{t\in [T]}{Y}_{ijt}.\]
The following lemma and corollary imply the identifiability of the
parameters.

\begin{lemma}\label{ThE}
Under Conditions \ref{conC}--\ref{conE}, we have
\begin{eqnarray}
\nonumber
\mathbb{E}(Y_{ijt}|\bm{Z}_{ijt})= (\theta_{i}^{\ast}-\theta_{j}^{\ast}+\bm{Z}_{ijt}^{\top}\bm{\eta}^{\ast})/\gamma_0^{\ast}.
\end{eqnarray}
\end{lemma}

The proof of Lemma \ref{ThE} is provided in the Supplementary Material.
Without loss of generality, we set $\gamma_0^{\ast}=1$ for addressing
the scale invariance issue. Alternatively, we could also use $\gamma_0^{\ast}=-1$
as the restricted condition.
Under the restrication $\gamma_0^{\ast}=1$,  we have
\[
\mathbb{E}(\overline{\bm{Y}}|\bm{Z}) =\bm{U}\bm{\theta}^{\ast}+ \overline{\bm{Z}}\bm{\eta}^{\ast},
\]
where $\bm{\theta}^*=(\theta_1^*,\ldots,\theta_n^*)^{\top}$ is the vector of the true values for $\bm{\theta}$.

Lemma \ref{ThE} suggests that if the model is identifiable, we can estimate the unknown parameters using $\overline{\bm{Y}}$, which implies that following corollary.

\begin{corollary}\label{Co1}
Under Conditions \ref{conC}--\ref{conU}, if $\gamma_{0}^{\ast}=1$ and $\theta_{0}^{\ast}=0$, we have
\begin{eqnarray}
\nonumber
\bm{\theta}^{\ast}
& = &\bm{V}^{-1}\bm{U}^{\top}\mathbb{E}(\overline{\bm{Y}} -\overline{\bm{Z}}\bm{\eta}^{\ast}),
\\  \nonumber
\bm{\eta}^{\ast} & = & \mathbb{E}(\overline{\bm{Z}}^{\top}\bm{D}\overline{\bm{Z}})^{-1} \mathbb{E}(\overline{\bm{Z}}^{\top}\bm{D}\overline{\bm{Y}}),
\end{eqnarray}
where $\bm{\theta}^*=(\theta_1^*,\ldots,\theta_n^*)^{\top}$ excludes $\theta_0^*$.
\end{corollary}

Corollary \ref{Co1} establishes the identifiability of the parameters.
Additionally, it suggests that we can estimate the unknown parameters using a least squares method.
Specifically,
let $\hat{f}(x|\bm{Z}_{ijt})$ be an estimator of $f(x|\bm{Z}_{ijt})$,
which is provided below.
Define
\begin{eqnarray}\label{yhat}
\widehat{Y}_{ijt}=\frac{a_{ijt}-\mathbb{I}(X_{ijt,0}>0)} {\hat{f}(X_{ijt,0}|\bm{Z}_{ijt})}.
\end{eqnarray}
Further, denote
\begin{eqnarray}\label{ybreve}
\breve{\bm{Y}}=(\breve{Y}_{01},\ldots,\breve{Y}_{0n},\breve{Y}_{12},\ldots,\breve{Y}_{1n},\ldots,\breve{Y}_{(n-1)n})^{\top} , ~~  \breve{Y}_{ij}=\frac{1}{T}\sum_{t\in [T]}\widehat{Y}_{ijt}.
\end{eqnarray}
By Corollary \ref{Co1}, we can estimate $\bm{\eta}^{*}$ and $\bm{\theta}^*$, respectively, using
\begin{equation}\label{betahat}
\begin{aligned}
\widehat{\bm{\eta}}=&(\overline{\bm{Z}}^{\top}\bm{D}\overline{\bm{Z}})^{-1} \overline{\bm{Z}}^{\top}\bm{D}\breve{\bm{Y}},\\
\text{and}~~\widehat{\bm{\theta}}=&\bm{V}^{-1}\bm{U}^{\top} [\bm{I}_{N}-\overline{\bm{Z}}(\overline{\bm{Z}}^{\top}\bm{D}\overline{\bm{Z}})^{-1} \overline{\bm{Z}}^{\top}\bm{D}]\breve{\bm{Y}}
= \bm{V}^{-1}\bm{U}^{\top} (\breve{\bm{Y}}-\overline{\bm{Z}}\widehat{\bm{\eta}}).
\end{aligned}
\end{equation}
Indeed, $\widehat{\bm{\eta}}$ and $\widehat{\bm{\theta}}$ are the following least squares estimators:
\begin{eqnarray}\label{etahat}
(\widehat{\bm{\theta}},\widehat{\bm{\eta}}) =\arg \min_{\bm{\theta},\bm{\eta}} \mathcal{L}(\bm{\theta},\bm{\eta}),
\end{eqnarray}
where
\[
\mathcal{L}(\bm{\theta},\bm{\eta})= \| \breve{\bm{Y}}-(\bm{U}\bm{\theta}+\overline{\bm{Z}}\bm{\eta}) \|_2^2.
\]

The computational complexity of the estimators $\widehat{\bm{\theta}}$ and $\widehat{\bm{\eta}}$
mainly involves algebraic operations on $\bm{Z}$, $\breve{\bm{Y}}$ and $\bm{D}$.
Note that $\breve{\bm{Y}}$ is given in \eqref{ybreve}, which depends on the computation of nonparametric  density estimator $\hat{f}$.
Its computational complexity is $O(n^2) \cdot O(n^2)=O(n^4)$ since
it requires evaluating $O(n^2)$ terms $\hat{f}_{ijt}(x_{ijt}|\bm{Z})$ and
the computational complexity of  each evaluation of $\hat{f}_{ijt}(x_{ijt}|\bm{Z})$ is $O(n^2)$.
Note that $\bm{\overline{Z}}$ is an $N\times p$ dense matrix with fixed $p$,
and $\bm{D}$ is a sparse matrix, where
each row and each column contain at least $O(n)$ nonzero elements.
Therefore, the computational complexity of $\overline{\bm{Z}}^{\top}\bm{D}$ is $O(n^3)$
and the computational complexity of $\overline{\bm{Z}}^{\top}\bm{D}\overline{\bm{Z}}$
is $O(n^3)$.
The computational complexity of $\overline{\bm{Z}}^{\top}\bm{D}\breve{\bm{Y}}$
for a given $\breve{\bm{Y}}$
is $O(n^3)$.
The total computational complexity of $\widehat{\bm{\eta}}$ is
$O(n^4)$.
Similarly, the total computational complexity of $\widehat{\bm{\theta}}$ is $O(n^4)$.

We next
employ the Nadaraya-Watson type estimator \citep{1964On, 1964Smooth} for the conditional density $f(x|\bm{z})$:
\begin{eqnarray}
\label{cfhat}
\hat{f}(x|\bm{z})
=
\frac{\sum_{0 \leq i\neq j \leq n} \sum_{t\in [T]} \mathcal{K}_{xz_1,h}(X_{ijt,0}-x,\bm{Z}_{ijt,1} -\bm{z}_1)\mathbb{I}(\bm{Z}_{ijt,2} =\bm{z}_2)} {\sum_{0 \leq i\neq j \leq n}\sum_{t\in [T]}  \mathcal{K}_{z_1,h}(\bm{Z}_{ijt,1} -\bm{z}_1)\mathbb{I}(\bm{Z}_{ijt,2}=\bm{z}_2)},~~~
\end{eqnarray}
where $\bm{Z}_{ijt,1}$ and  $\bm{Z}_{ijt,2}$ denote the continuous covariates and the discrete covariates in $\bm{Z}_{ijt}$, respectively. Here,
\[
\mathcal{K}_{z_1,h}(\bm{Z}_{ijt,1} -\bm{z}_1)
=\frac{1}{h^{p_1}}\mathcal{K}_{z_1}\left(\frac{\bm{Z}_{ijt,1} -\bm{z}_1}{h}\right),\]
\[
\mathcal{K}_{xz_1,h}(X_{ijt,0}-x,\bm{Z}_{ijt,1} -\bm{z}_1)
=\frac{1}{h^{p_1+1}}\mathcal{K}_{xz_1}\left(\frac{X_{ijt,0}-x}{h},\frac{\bm{Z}_{ijt,1} -\bm{z}_1}{h}\right),\]
where $h$ denotes the bandwidth parameter, $p_1$ denotes the dimension of the continuous covariates $\bm{Z}_{ijt,1}$, and $\mathcal{K}_{z_1}(\cdot)$ and $\mathcal{K}_{xz_1}(\cdot)$ are two kernel functions.

\section{Theoretical Results}\label{Sec3}
\renewcommand{\theequation}{3.\arabic{equation}}
\setcounter{equation}{0}
In this section, we establish consistency and asymptotic normality for the estimators. We consider the following conditions.

\begin{condition}\label{conZ}
For all $1\le i\neq j\le n$ and $t\in[T]$,
$\|\bm{Z}_{ijt}\|_{\infty}\leq \kappa$ almost surely,
where $\kappa$ is allowed to diverge with $n$.
\end{condition}
\begin{condition}\label{conf}
The density functions satisfy $\min\{f(x_{ijt,0}|\bm{Z}_{ijt}),f(\bm{z}_{ijt})\}>\varpi>0$ on the support of $X_{ijt,0}$ almost surely.
In addition, the $r$th order partial derivative of the density function $f(\bm{z}_1)$ of $\bm{Z}_{ijt,1}$ exists and is continuous and bounded.
The $r$th order partial derivative of the joint density function $f(x, \bm{z}_1)$ of $(X_{ijt,0}, \bm{Z}_{ijt,1})$ is also continuous and bounded.
Here, $\varpi$ is allowed to decrease to zero as $n \rightarrow \infty$, and the subscript $n$ in $\varpi$ is suppressed.
\end{condition}
\begin{condition}\label{conK}
The kernel function $\mathcal{K}_{\bm{z}}(\bm{z})$ is symmetric and piecewise Lipschitz continuous with order $r$.
That is,
\begin{equation*}
\begin{aligned}
\int\cdots\int \mathcal{K}_{\bm{z}}(z_1,\ldots,z_{p_1})dz_1\cdots dz_{p_1}&=1,\\
\int\cdots\int z_1^{r_1}\cdots z_{p_1}^{r_{p_1}} \mathcal{K}_{\bm{z}}(z_1,\ldots,z_{p_1})dz_1\cdots dz_{p_1}&=0, ~~(0<r_1+\cdots+r_{p_1}<r),\\
\int\cdots\int z_1^{r_1}\cdots z_{p_1}^{r_{p_1}} \mathcal{K}_{\bm{z}}(z_1,\ldots,z_{p_1})dz_1\cdots dz_{p_1}& \neq 0, ~~(0<r_1+\cdots+r_{p_1}=r),
\end{aligned}
\end{equation*}
where $p_1$ denotes the dimension of $\bm{Z}_{ijt,1}$.
\end{condition}

Condition \ref{conZ} assumes that the covariates are bounded above.
This assumption significantly simplifies the proofs of the following theorems.
However, it can be relaxed to sub-Gaussian covariates.
Conditions \ref{conf} and \ref{conK} are mild and widely used in nonparametric kernel smoothing methods.

Combining Lemma \ref{ThE}, the definition of $Y_{ijt}$ in \eqref{Y} and Condition \ref{conf},  we have
\[|\mathbb{E}(Y_{ijt})|= |\theta_{i}^{\ast}-\theta_{j}^{\ast}+\mathbb{E}(\bm{Z}_{ijt}^{\top})\bm{\eta}^{\ast}|
=
\left|\mathbb{E}\left(\frac{a_{ijt}-\mathbb{I}(X_{ijt,0}>0)}{f(X_{ijt,0}|\bm{Z}_{ijt})}\right)\right|
\leq O\left(\frac{1}{\varpi}\right).\]
This implies that $|\theta_{i}^{\ast}-\theta_{j}^{\ast}|$ is bounded above by $O(1/\varpi)$.

The following theorem establishes the consistency of the estimators for the merit parameters and the regression coefficients.

\begin{theorem}\label{TheC}
Suppose that Conditions \ref{conC}--\ref{conK} hold. If
\begin{eqnarray}\label{O1}
 \frac{1}{\varpi^2}\left(1+\frac{\kappa^2}{\lambda}\right) \left(\sqrt{\frac{\log n}{n }}+\frac{\sqrt{\log n}}{nh^{p+1}}+h^r\right)=o(1),
\end{eqnarray}
then we have
\[
\|\widehat{\bm{\eta}}-\bm{\eta}^{\ast}\|_{\infty}=o_p(1), \quad
\|\widehat{\bm{\theta}}-\bm{\theta}^{\ast}\|_{\infty}=o_p(1).
\]
\end{theorem}

Condition \eqref{O1} involves the selection of bandwidth $h$ to balance the bias and variance of $\widehat{Y}_{ijt}$ when using the kernel smoothing method.
Thus, the selection of bandwidth $h$ influences the performance of the estimator
both theoretically and practically. When $\kappa/\varpi$ is of a constant order,
it is necessary for $h\rightarrow 0$ and $nh^{p+1}/ \sqrt{\log n} \rightarrow \infty$ as $n \rightarrow \infty$ to ensure the consistency of the
estimators.

We now state the asymptotic distribution of the kernel-based least squares estimator.
For convenience, define
\begin{eqnarray}\label{tau}
\overline{\bm{\tau}}=(\overline{\tau}_{01},\ldots,\overline{\tau}_{0n},
\overline{\tau}_{12},\ldots,\overline{\tau}_{1n},\ldots,\overline{\tau}_{(n-1)n})^{\top}
:= \overline{\bm{Y}}-\mathbb{E}(\overline{\bm{Y}}|\bm{X}_{0},\bm{Z}),
\\
\label{xi}
\overline{\bm{\xi}}=(\overline{\xi}_{01},\ldots,\overline{\xi}_{0n},
\overline{\xi}_{12},\ldots,\overline{\xi}_{1n},\ldots,\overline{\xi}_{(n-1)n})
^{\top}
:=\overline{\bm{Y}}-\mathbb{E}(\overline{\bm{Y}}|\bm{Z}).
\end{eqnarray}
Since all elements of $\overline{\bm{\tau}}$ are independent, the covariance matrix of $\overline{\bm{\tau}}$ can be written as
\begin{equation}
\label{definition-Sigma-tau}
\bm{\Sigma}_{\tau}=\diag(\sigma^2_{\tau,01},\ldots,\sigma^2_{\tau,0n},\sigma^2_{\tau,12},\ldots,\sigma^2_{\tau,1n},\ldots,\sigma^2_{\tau,(n-1)n})
:=\mathrm{Cov}( \overline{\bm{\tau}} ),
\end{equation}
where $\sigma^2_{\tau,ij}:= \mathrm{Var}(\overline{\tau}_{ij})$.
Similarly, we define the covariance matrix of $\overline{\bm{\xi}}$ as
\begin{equation}
\label{definition-Sigma-xi}
\bm{\Sigma}_{\xi}= \diag(\sigma^2_{\xi,01},\ldots,\sigma^2_{\xi,0n},\sigma^2_{\xi,12},\ldots,\sigma^2_{\xi,1n},\ldots,\sigma^2_{\xi,(n-1)n}).
\end{equation}
The asymptotic normality of $\widehat{\bm{\eta}}$ is stated below.

\begin{theorem}
\label{TheCLTe}
Suppose that Conditions \ref{conC}--\ref{conK} hold.
If $\sup_{i,j} \sigma^2_{\tau,ij}<\infty$ and
\begin{eqnarray}\label{O2}
\frac{\kappa}{\lambda \varpi}\left(\frac{\sqrt{\log n}} {nh^{p+1}} +nh^r \right)=o(1) \text{ as }n \rightarrow \infty,
\end{eqnarray}
then $n (\widehat{\bm{\eta}}-\bm{\eta}^*)$ is
asymptotically normally distributed with mean $\bm{0}$ and covariance matrix
$n^2 \mathbb{E}[(\overline{\bm{Z}}^{\top}\bm{D}\overline{\bm{Z}})^{-1}] \mathbb{E}(\overline{\bm{Z}}^{\top}\bm{D}\bm{\Sigma}_{\tau}\bm{D}\overline{\bm{Z}})
\mathbb{E}[(\overline{\bm{Z}}^{\top}\bm{D}\overline{\bm{Z}})^{-1}]$,
where $\bm{\Sigma}_{\tau}$ is defined in \eqref{definition-Sigma-tau}.
\end{theorem}

When $\sigma^2_{\tau,ij}=\sigma^2_{\tau}$ for all $i\neq j$, we have
 $n (\widehat{\bm{\eta}}-\bm{\eta}^*)$
is asymptotically normally distributed with mean $\bm{0}$ and variance matrix $n^2 \sigma_{\tau}^2 \mathbb{E}[(\overline{\bm{Z}}^{\top}\bm{D}\overline{\bm{Z}})^{-1}]$ by noting that
\begin{eqnarray*}
&&\mathbb{E}[(\overline{\bm{Z}}^{\top}\bm{D}\overline{\bm{Z}})^{-1}] \mathbb{E}(\overline{\bm{Z}}^{\top}\bm{D}\bm{\Sigma}_{\tau}\bm{D}\overline{\bm{Z}})
\mathbb{E}[(\overline{\bm{Z}}^{\top}\bm{D}\overline{\bm{Z}})^{-1}]
\\ =&& \sigma_{\tau}^2\mathbb{E}[(\overline{\bm{Z}}^{\top}\bm{D}\overline{\bm{Z}})^{-1}] \mathbb{E}(\overline{\bm{Z}}^{\top}\bm{D}\bm{D}\overline{\bm{Z}})
\mathbb{E}[(\overline{\bm{Z}}^{\top}\bm{D}\overline{\bm{Z}})^{-1}]
= \sigma_{\tau}^2 \mathbb{E}[(\overline{\bm{Z}}^{\top}\bm{D}\overline{\bm{Z}})^{-1}] .
\end{eqnarray*}

The asymptotic distribution of $\widehat{\bm{\theta}}$ is stated below.

\begin{theorem}
\label{TheCLTb}
Suppose that Conditions \ref{conC}--\ref{conK} hold.
If  $\sup_{i,j} \sigma^2_{\xi,ij}:= \mathrm{Var}(\overline{\xi}_{ij})<\infty$ and
\begin{eqnarray}\label{O3}
\frac{1}{\varpi^2} \left(1+\frac{\kappa^2}{\lambda}\right) \left(\sqrt{\frac{\log n}{n}}+
\sqrt{\frac{\log n}{nh^{2p+2}}}+\sqrt{n}h^r\right)=o(1)
\text{ as }n \rightarrow \infty,~~~
\end{eqnarray}
then, for a constant vector $\bm{c}=(c_1,\ldots,c_n)^{\top}$ satisfying $\sum_{i=1}^{n}|c_i|<\infty$, we have
$\sqrt{n}\bm{c}^{\top}(\widehat{\bm{\theta}}-\bm{\theta}^{\ast})$
is asymptotically normally distributed with mean $0$ and variance
$n\bm{c}^{\top}\bm{V}^{-1}\bm{U}^{\top}
\bm{\Sigma}_{\xi} \bm{U}\bm{V}^{-1}\bm{c}$,
where $\bm{\Sigma}_{\xi}$ is defined in \eqref{definition-Sigma-xi}.
\end{theorem}

When $\sigma^2_{\xi,ij}=\sigma^2_{\xi}$ for all $i\neq j$, we have that
$\sqrt{n}\bm{c}^{\top}(\widehat{\bm{\theta}}-\bm{\theta}^{\ast})$
is asymptotically normally distributed with mean $0$ and variance $n\bm{c}^{\top}\bm{V}^{-1}\bm{c}$ by noting that
\begin{eqnarray*}
\bm{c}^{\top}\bm{V}^{-1}\bm{U}^{\top}\bm{\Sigma}_{\xi}
 \bm{U}\bm{V}^{-1}\bm{c}
=
\bm{c}^{\top}\bm{V}^{-1}\bm{V}\bm{V}^{-1}\bm{c}
=\bm{c}^{\top}\bm{V}^{-1}\bm{c}.
\end{eqnarray*}

\section{Numerical studies}\label{Sec4}
\renewcommand{\theequation}{4.\arabic{equation}}
\setcounter{equation}{0}

We first consider a data-driven procedure to determine the bandwidth $h$.
Specifically, let $\delta$ be an arbitrarily given constant.  It is easy to verify that
\[
\delta=\mathbb{E}\{[\mathbb{I}(X_{ijt,0}+\delta >0)-\mathbb{I}(X_{ijt,0} >0)]/f(X_{ijt,0} |\bm{Z}_{ijt} )\}.
\]
Define
$$\hat{\delta}(h)=\frac{1}{NT}\sum_{0\leq i\neq j\leq n}\sum_{t\in[T]}\frac{\mathbb{I}(X_{ijt,0}+\delta >0)
-\mathbb{I}(X_{ijt,0} >0) }{\hat{f}(X_{ijt,0} |\bm{Z}_{ijt} )}.$$
Motivated by \cite{lewbel1998semiparametric}, we can estimate $h$ by
\[
\hat{h}=\arg\min_{h} \sum_{m=1}^M[\delta_m-\hat{\delta}_m(h)]^2,
\]
where $\delta_m\,(1\leq m\leq M)$ are pre-specified grid points on $(0,1]$, and $M$ is a pre-specified integer.
In the simulation studies, we set $\delta_{m}\in\{0.1,0.2,\dots,0.9\}$.
For the kernel function, we employ the quartic kernel function
\begin{eqnarray}\label{Ker}
\mathcal{K}_{z}(\bm{z})=\prod_{l} \frac{15}{16} (1-z_l^2)^2 \mathbb{I}(|z_l|\leq 1),
\end{eqnarray}
due to its computational efficiency and favorable properties; see \citet{hardle1990applied}. The quartic kernel function in \eqref{Ker} is symmetric and piecewise Lipschitz continuous with order two, satisfying Condition \ref{conK}.

According to the definition of $\hat{h}$, its computational complexity depends on the number of candidate bandwidths $k_h$ $(h_1, \ldots, h_{k_h})$, the number of pre-specified grid points $\delta_m\ (1\leq m\leq M)$ on $(0,1]$, and the computation of the nonparametric density estimator $\hat{f}(X_{ijt,0} |\bm{Z}_{ijt} )$.
Because the dimension of $\bm{Z}_{ijt}$ is fixed,
the computational complexity of $\hat{f}(X_{ijt,0} |\bm{Z}_{ijt} )$ is $O(n^2)$.
This leads to the $O(n^4)$-computational complexity of $\hat{\delta}(h)$.
As a result, the total computational complexity for computing $\hat{h}$ is $O(k_h Mn^4)$.

\subsection{Simulation studies}
\label{Simulation}

In this section, we evaluate the finite sample performance of the proposed method. We generate the covariates $\bm{Z}_{ijt}$ from a two-dimensional normal distribution with mean zero
and covariance $\bm{\Sigma}=( \sigma_{ij} )_{2\times 2}$.
Here, we take $\sigma_{11}=\sigma_{22}=1$ and $\sigma_{12}=\sigma_{21}=1/4$.
To construct a special regressor $\bm{X}_0$
that depends on covariates
$\bm{Z}$ but does not affect their values, we set $X_{ijt,0}=\bm{Z}_{ijt}^{\top}\bm{b} + \omega_{ijt}$,
where $\bm{b}=(0.5,-0.5)^{\top}$, and $\omega_{ijt}$ is independently generated from the standard normal distribution.
The parameter $\theta_i^{\ast}$ is set to $\theta_i^{\ast}=0.2i\log n /n$.
We set $\bm{\eta}^{\ast}=(-0.5,0.5)^{\top}$.
For the noise term $\varepsilon_{ijt}$, we consider the following three cases:
\begin{itemize}
\item[(i)] $\varepsilon_{ijt}$ is generated from the standard normal distribution $N(0, 1)$.

\item[(ii)] $\varepsilon_{ijt}$ is generated from the logistic distribution
with shift parameter $0$ and scale parameter $\sqrt{3}/\pi$, i.e.,
 $\varepsilon_{ijt}\sim \text{Logistic}(0, \sqrt{3}/\pi)$, where $\mathrm{Var}(\varepsilon_{ijt})=1$.
Here, $\text{Logistic}(\mu, s)$ denotes the logistic distribution with
the density function
\begin{equation}
\label{eq-logis-density}
f(x; \mu, s)= \frac{\exp \left(-\frac{x-\mu}{s}\right)} {s\left(1+\exp \left(-\frac{x-\mu}{s}\right)\right)^2},
\end{equation}
where $\mu$ is the shift parameter (controlling the center of the distribution), and $s > 0$ is the scale parameter (controlling the dispersion of the distribution).

\item[(iii)] $\varepsilon_{ijt}$ is generated from $N(-0.3, 0.91)$ with probability $0.75$ and $N(0.9, 0.19)$
with probability $0.25$, denoted as $\text{Mix-Norm}$.
\end{itemize}
The first case corresponds to the probit regression model, while the second case involves a logistic regression model.
The last case  concerns a mixture of normal distributions,
designed to produce a distribution that is skewed and bimodal,
while maintaining mean zero and variance one.

Each simulation is repeated $1000$ times.
The average biases, standard deviations (SD), and coverage frequencies (CP) of the $95\%$ confidence intervals
for the estimate $\widehat{\theta}_i$ and $\widehat{\eta}_i$ ($i=1,2$) are recorded.
Here, we present the results for $\widehat{\theta}_i$ with $i=\{1, 12, 25, 37, 50\}  \text{ for } n=50 \text{ and } i=\{1, 25, 50, 75, 100\}  \text{ for } n=100$.
Table \ref{tab1} reports the results for Case (i),
while the results for Cases (ii) and (iii) are reported in 
the Supplementary Material.

Table \ref{tab1} demonstrates the good performance of the proposed method.
Specifically, the bias of the proposed estimators is very small, and the simulated coverage frequencies
are very close to the $95\%$ target level.
As expected, this table shows that the bias decreases as the number of items $n$ increases.
The standard deviation also decreases as $n$ increases.
When the number of paired comparisons $T$ for each pair increases, the bias and the standard deviation decrease.
Similar phenomena for Cases (ii) and (iii) are also observed
in the Supplementary Material.
\begin{table}[ht]\centering
\caption{Simulation results when $\varepsilon_{ijt}\sim N(0,1)$.}\label{tab1}
\footnotesize
\vspace{0.2cm}
\begin{tabular}{cccccccccc}
\hline
 & &&\multicolumn{3}{c}{$T=1$}
&&\multicolumn{3}{c}{$T=3$}\\
\cmidrule(r){4-6}\cmidrule(r){8-10}
$n$& &&Bias &SD  &CP
&&Bias &SD  &CP\\
\midrule
50
&$\theta_1$	&&	~~0.0054 	&	0.348 	&	0.939 	&&	~~0.0042 	&	0.210 	&	0.955 	\\
&$\theta_{12}$	&&	$-$0.0019 	&	0.353 	&	0.948 	&&	~~0.0090 	&	0.208 	&	0.952 	\\
&$\theta_{25}$	&&	$-$0.0113 	&	0.350 	&	0.948 	&&	~~0.0047 	&	0.204 	&	0.945 	\\
&$\theta_{37}$	&&	$-$0.0066 	&	0.361 	&	0.951 	&&	$-$0.0106 	&	0.207 	&	0.954 	\\
&$\theta_{50}$	&&	$-$0.0178 	&	0.367 	&	0.964 	&&	$-$0.0113 	&	0.206 	&	0.955 	\\
&$\eta_1$	&&	$-$0.0053 	&	0.047 	&	0.947 	&&	$-$0.0017 	&	0.027 	&	0.954 	\\
&$\eta_2$	&&	~~0.0026 	&	0.047 	&	0.949 	&&	~~0.0033 	&	0.027 	&	0.949 	\\
\midrule	
100			
&$\theta_1$	&&	$-$0.0025 	&	0.283 	&	0.950 	&&	$-$0.0019 	&	0.167 	&	0.956 	\\
&$\theta_{25}$	&&	~~0.0094 	&	0.280 	&	0.959 	&&	~~0.0067 	&	0.156 	&	0.955 	\\
&$\theta_{50}$	&&	~~0.0025 	&	0.283 	&	0.958 	&&	$-$0.0001 	&	0.156 	&	0.950 	\\
&$\theta_{75}$	&&	$-$0.0022 	&	0.282 	&	0.956 	&&	$-$0.0054 	&	0.150 	&	0.958 	\\
&$\theta_{100}$	&&	$-$0.0081 	&	0.276 	&	0.967 	&&	$-$0.0065 	&	0.158 	&	0.947 	\\
&$\eta_1$	&&	$-$0.0025 	&	0.025 	&	0.954 	&&	$-$0.0037 	&	0.014 	&	0.934 	\\
&$\eta_2$	&&	~~0.0037 	&	0.025 	&	0.947 	&&	~~0.0028 	&	0.015 	&	0.949 	\\
\bottomrule
\end{tabular}
\end{table}

To assess the asymptotic normality of the parameter estimators, we provide QQ plots (quantile-quantile plots) of the kernel-based least squares estimators.  The QQ plots of $\eta_1$ and $\theta_1$ for Cases (i)--(iii) are reported in 
the Supplementary Material. Most of the data points of the estimators from our proposed method fall near the reference lines, indicating that when the sample size is sufficiently large, the sampling distributions of both
$\widehat{\eta}_1$ and $\widehat{\theta}_1$
approximately follow normal distributions. However, there is a slight deviation of $\theta_1$ in the tails, which may indicate a mildly heave-tailed behavior. In comparison, $\eta_1$ fits better, indicating that its asymptotic normality is more robust.

We have conducted simulation studies for a comprehensive comparison
between the semiparametric model with the parametric model (i.e., the covariate-adjusted Bradley--Terry model \citep{yan2020paired}).
Specifically, we consider the covariate-adjusted
Bradley--Terry model as follows:
\begin{equation}
\label{model-yan-c1}
\mathbb{P}(A_{ijt}=1|X_{ijt,0},\bm{Z}_{ijt})
=\frac{\exp(\theta_{i}-\theta_{j}+X_{ijt,0}+\bm{Z}_{ijt}^\top\bm{\eta})}
{1+\exp(\theta_{i}-\theta_{j}+X_{ijt,0}+\bm{Z}_{ijt}^\top\bm{\eta})},
\end{equation}
where $\theta_0$ is set to be $0$ and the coefficient of $X_{ijt,0}$  is fixed at $1$ for a fair comparison.
Note that \cite{fan2024uncertainty} assumed that the covariates of each item
enter into the Bradley--Terry model with an additive form,  where
item $i$ beats item $j$ with probability $e^{\mu_{ij}}/(1 + e^{\mu_{ij}})$ with
 $\mu_{ij}=\theta_i -\theta_i + X_i^\top \gamma
- X_j^\top \gamma$. Here, $X_i$ denotes the covariate of item $i$.
It can be viewed as a special case of the model in \cite{yan2020paired}, where
the pairwise covariates associated with each comparison are used.
We use the maximum likelihood to estimate the unknown parameters
when fitting the above model to the simulated data.

When model \eqref{model-yan-c1} is correctly specified,
 the noise in model \eqref{Model} follows a standard logistic distribution, denoted as
 $\mathrm{Logistic}(0,1)$.
We consider four scenarios for the noise $\epsilon_{ijt}$:
(1) $\epsilon\sim N(0,1)$, (2) $\epsilon\sim \text{Logistic}(0, \sqrt{3}/\pi)$, (3) $\epsilon \sim \text{Mix-Norm}$
and (4) $\epsilon \sim \text{Logistic}(0,1)$,
where the density of a logistic distribution is given in \eqref{eq-logis-density} and
  $\text{Mix-Norm}$ denotes a mixture
of normal distributions that is the same as case (iii) in the simulation.
We set the merit parameters $\theta_i^{\ast}$ as $\theta_i^{\ast}=0.2i\log n /n$
and the covariate parameter $\bm{\eta}^{\ast}=(-0.5,0.5)^{\top}$.
Each simulation was repeatedly 1000 times.
We compare the biases of the estimator obtained from the semiparametric model with that of
the MLE in the covariate-adjusted Bradley--Terry model.
The results are presented in Table \ref{tabMLE1}.

 \begin{table}[ht]\centering
\caption{The comparison results of the bias of the estimators obtained
by our method and the MLE method with $n=100$ and $T = 1$.}\label{tabMLE1}
\tiny
\vspace{0.2cm}
\begin{tabular}{cccccccccccc}
\hline
  &  \multicolumn{2}{c}{$N(0,1)$}
&&\multicolumn{2}{c}{$\text{Logistic}(0,\sqrt{3}/\pi)$}
&&\multicolumn{2}{c}{$\text{Mix-Norm}$}
&&\multicolumn{2}{c}{$\text{Logistic}(0,1)$}\\
\cmidrule(r){2-3} \cmidrule(r){5-6} \cmidrule(r){8-9} \cmidrule(r){11-12}
 &Our method & MLE &&Our method & MLE
&&Our method & MLE &&Our method & MLE \\
\midrule															
	$\theta_1$	&	$-$0.0025 	&	0.0009 	&&	$-$0.0093 	&	$-$0.0024 	&&	~~0.0115 	&	~~0.0083 	&&	$-$0.0054 	&	$-$0.0090 	\\
	$\theta_{25}$	&	~~0.0025 	&	0.0880 	&&	$-$0.0099 	&	~~0.0972 	&&	~~0.0065 	&	~~0.1145 	&&	$-$0.0448 	&	$-$0.0123 	\\
	$\theta_{50}$	&	$-$0.0081 	&	0.1672 	&&	$-$0.0160 	&	~~0.2059 	&&	~~0.0061 	&	~~0.2159 	&&	$-$0.0699 	&	~~0.0122 	\\
    $\eta_{1}$	&	$-$0.0025 	&	0.0009 	&&	$-$0.0023 	&	$-$0.0005 	&&	$-$0.0044 	&	$-$0.0006 	&&	$-$0.0259 	&	~~0.0000 	\\
	$\eta_{2}$	&	~~0.0037 	&	0.0001 	&&	~~0.0030 	&	~~0.0014 	&&	~~0.0050 	&	~~0.0001 	&&	~~0.0281 	&	~~0.0014 	\\
\bottomrule
\end{tabular}
\end{table}

From Table \ref{tabMLE1}, we observe that when the covariate-adjusted Bradley--Terry model is correctly specified
(i.e., the noise is distributed according to a standard logistic distribution), in most cases,
the MLE has smaller biases than the estimator obtained from the semiparametric model.
This is expected since parametric models  use more information and are more efficient than
semiparametric models when they are correctly specified.
When the covariate-adjusted Bradley--Terry model is not correctly specified
(i.e., the noise is not generated from the standard logistic distribution),
the semiparametric estimator has smaller biases than the MLE.
For instance, when $\epsilon \sim N(0,1)$ or $\epsilon\sim \mathrm{Logistic}(0, \sqrt{3/\pi})$,
the biases of the semiparametric estimator are close to zero, whereas the bias of the MLE for parameter $\theta_{50}$ in the misspecified model \eqref{model-yan-c1} exceeds  $0.16$.
Additionally, we note that the MLE for the covariate parameters remains unbiased,
even when the noise distribution is misspecified.
This phenomenon may arise because $n^2$ samples are used to estimate the fixed-dimensional covariate parameters,
thereby effectively attenuating the influence of the misspecified noise distribution on the MLE of the covariate parameters.

\subsection{Real data analysis}\label{secNBA}
In this section, we analyze a real dataset using the proposed method.
We consider an American National Basketball Association (NBA) dataset for
the season 2018-2019. The data are freely available from the Basketball
Excel website (\url{https://www.basketball-reference.com}), which provides detailed information about each season, such as the result of each game and the winning percentage for each team.
The NBA has 30 teams. The league is divided into Eastern and Western conferences, with 15 teams in each conference.
From October 2018 to April 2019, each team plays 82 games, 41 at home and 41 away. The total number of matches is 1230. We focus on the results of the games, which have a binary outcome: $a_{ijt} = 1$ when team $i$ defeats team $j$ in the $t$-th match; otherwise, $a_{ijt} = 0$.

As argued in \citet{Cat13} and \citet{Tut15}, home-field advantage influences  sports matches, because
the home team has obvious advantages due to familiarity with the competition environment and support from the home crowd. Conversely, the away team faces the challenge of traveling and adapting to an unfamiliar environment.
Let $Z_{ijt,1}$ be the covariate indicating home-field advantage  information.
If the home team $i$ plays with the visiting team $j$ in the $t$th match, then $Z_{ijt,1}=1$ and $Z_{jit,1}=-1$.
Additionally, \citet{2020Basketball} highlighted the importance of taking at least one day of rest to increase the chance of winning a match, demonstrating the detrimental effect of back-to-back matches on game outcomes. In the NBA, ``back-to-back" specifically denotes two successive days of away games. Typically, teams experiencing highly congested schedules may be particularly exposed to air travel effects, circadian rhythm disruption, sleep deprivation, decline in physical capacity, and injury risk ~\citep{Reilly01}.
Consequently, we regard back-to-back matches as a covariate in our analysis to accurately determine if the away team is at a competitive disadvantage due to consecutive matches.
Let $Z_{ijt,2}$ be the covariate indicating whether the away team plays back-to-back games.
If team $i$ is the away team in a back-to-back game and team $j$ is the home team,
then  $Z_{ijt,2}=1$ and $Z_{jit,2}=-1$; otherwise, $Z_{ijt,2}=Z_{jit,2}=0$.

We now define a special regressor under Condition \ref{conC}. We employ the win percentage as the continuous covariate, and ensure its timeliness through dynamic adjustment. In the first month of the season, due to the lack of actual data for that season, we calculate the predicted winning percentage using the win percentagwe calculate the predicted winning percentage using the predicted wins released on October 16, 2018 by \href{https://www.espn.com/nba/insider/story/_/id/24988536/projected-nba-win-totals-playoff-standings-all-30-teams-2018-19}{ESPN}'s well-known analyst Kevin Pelton, whose methodology incorporates changes in team rosters, player health, and outlook for the season.
Starting from the second month, we utilize a rolling monthly net win percentages metric, calculated as the number of wins of the team divided by the total number of games the team played in the previous month. This hybrid method balances between pre-season analytics and in-season performance. Specifically, let $X_{ijt,0}=W_{it,j}-W_{jt,i}$, where $W_{it,j}$ denotes the winning percentage of the  team $i$ at the $t$th match with $j$. We first partition the support of $X_{ijt,0}$ into $K=5$ equal subintervals, denoted by $\mathcal{I}_k=[x_k,x_{k+1})~(1\le k\le K)$, respectively.
We then calculate the wining rate when the value of $X_{ijt,0}$ falls within $\mathcal{I}_k$:
\begin{eqnarray}\label{WR}
\text{WR}_k=\frac{\sum_{t=1}^T\sum_{i=1}^n\sum_{j\neq i}a_{ijt,0}\mathbb{I}(X_{ijt,0}\in\mathcal{I}_k)}
{\sum_{t=1}^T\sum_{i=1}^n\sum_{j\neq i}\mathbb{I}(X_{ijt,0}\in\mathcal{I}_k)}.
\end{eqnarray}
The values of $\text{WR}_k$ are $0.864, 0.764, 0.613, 0.397$, and 0.130 for $k=1,\dots,5$, respectively,
exhibiting a decreasing trend as $k$ increases.
Therefore, we set $\gamma_0=-1$ for identifiablity.

We obtain the estimators according to \eqref{betahat}. For regression coefficients, $\widehat{\bm{\eta}}=(0.065,-0.098)^{\top}$, which indicates that home-field advantage positively affects the competition result with an intensity of $0.065$, whereas the back-to-back match negatively influences the competition result with an intensity of $0.098$. The $p$-value and a 95\% confidence interval (CI) for the estimated regression coefficient of the home-field advantage are $0.027$ and  $(0.008, 0.122)$, respectively. The $p$-value for the estimated regression coefficient of the back-to-back is  $0.234$.
This suggests that the home-field advantage has a significant positive effect on competition outcomes, while the back-to-back  has no significant effect on competition outcomes. We normalize the merit parameter of New York Knicks to $0$ for identifiability. New York Knicks has the  lowest  merit parameter estimator in the Eastern Conference, whereas Milwaukee Bucks has the highest with an estimator of $1.981$. The team has the highest estimated merit parameter in the Western Conference is Golden State Warriors, whereas the lowest is Phoenix Suns, with estimators of $1.874$ and $0.043$, respectively. Table \ref{tab4} shows the estimators of merit parameters, as well as $a_i$ (the number of wins for team $i$) and rank estimators $\widehat{R}_i$.
Due to the restriction of the page width, the $95\%$ confidence intervals of merit parameters are put in 
the Supplementary Material.

\begin{table}[ht]\centering
\caption{The estimates of $\theta_i$ in 2018-19 NBA regular season.}
\label{tab4}
\footnotesize
\vspace{0.2cm}
\begin{tabular}{ccccccccc}
\hline
\toprule
 \multicolumn{4}{c}{Eastern Conference}  &&\multicolumn{4}{c}{Western
Conference}\\
\cmidrule(r){1-4}\cmidrule(r){6-9}
Team	&$a_i$	&$\widehat{\theta}_i$	&$\widehat{R}_i$	
&&Team	&$a_i$	&$\widehat{\theta}_i$	&$\widehat{R}_i$	\\
\midrule
Milwaukee Bucks	&	60	&	1.981 	&	1	&&	Golden State Warriors	&	57	&	1.874 	&	3	\\
Toronto Raptors	&	58	&	1.978 	&	2	&&	Denver Nuggets	&	54	&	1.669 	&	5	\\
Philadelphia 76ers	&	51	&	1.646 	&	6	&&	Houston Rockets	&	53	&	1.680 	&	4	\\
Boston Celtics	&	49	&	1.608 	&	7	&&	Portland Trail Blazers	&	53	&	1.365 	&	11	\\
Indiana Pacers	&	48	&	1.579 	&	8	&&	Utah Jazz	&	50	&	1.394 	&	10	\\
Orlando Magic	&	42	&	0.879 	&	21	&&	Oklahoma City Thunder	&	49	&	1.491 	&	9	\\
Brooklyn Nets	&	42	&	1.044 	&	16	&&	San Antonio Spurs	&	48	&	1.323 	&	12	\\
Detroit Pistons	&	41	&	1.241 	&	14	&&	LA Clippers	&	48	&	1.243 	&	13	\\
Miami Heat	&	39	&	1.062 	&	15	&&	Sacramento Kings	&	39	&	0.880 	&	20	\\
Charlotte Hornets	&	39	&	0.905 	&	19	&&	Los Angeles Lakers	&	37	&	0.970 	&	17	\\
Washington Wizards	&	32	&	0.795 	&	23	&&	Minnesota Timberwolves	&	36	&	0.950 	&	18	\\
Atlanta Hawks	&	29	&	0.448 	&	26	&&	New Orleans Pelicans	&	33	&	0.824 	&	22	\\
Chicago Bulls	&	22	&	0.200 	&	27	&&	Dallas Mavericks	&	33	&	0.696 	&	25	\\
Cleveland Cavaliers	&	19	&	0.111 	&	28	&&	Memphis Grizzlies	&	33	&	0.713 	&	24	\\
New York Knicks	&	17	&	0 	&	30	&&	Phoenix Suns	&	19	&	0.043 	&	29	\\
\bottomrule
\end{tabular}
\end{table}

\section{Summary}
\label{Sec5}
This study introduced covariates that influence the outcomes of paired comparisons and constructed a semiparametric model for paired comparisons.
By introducing a special regressor, we developed a kernel-based least squares method to estimate all unknown parameters in the model.
The kernel-based least squares estimators have explicit expressions and can be extended to a large number of items.

When there are several continuous covariates that could be potentially used as the special regressor,
except for selecting the one that with the largest observed support,
we could loop over each special regressor to  estimate the model parameters and then take the average.
This model averaging approach can be used to estimate the merit parameters of items, but cannot be used to estimate the regression coefficient of the covariates because different special regressors lead to different regression coefficients of covariates.
However, this procedure brings difficulties for theoretical analyses because
the dependent structure of the estimators obtained from semiparametric models with different special regressors is very complex.
Another way is to use the cross-validation method based on some criterion (e.g., prediction error), but this method is very time-consuming in semiparametric paired comparison setting.

In Condition \ref{conC}, the proposed estimation procedure requires selecting a continuous covariate  $X_{ijt,k}$ whose corresponding regression coefficient is positive, and for convenience and identifiability, we normalize it to $1$. However, in empirical applications, we could also set the coefficient of the special regressor to be $-1$, and all theoretical results still hold. To
determine the sign of coefficient for the special regressor, we can partition
the support of covariate $X_{ijt,0}$ into $K$ equal subintervals, as described
in Section \ref{secNBA}.
If $\text{WR}_k$ in \eqref{WR} exhibits an increasing trend as $k$ increases, then the effect of the covariate $X_{ijt,0}$ is likely positive, and we set $\gamma_{0}^*=1$.
Conversely, if $\text{WR}_k$  shows a decreasing trend as $k$ increases, then the effect of the covariate $X_{ijt,0}$ is likely negative, and we set $\gamma_{0}^*=-1$.
However, when there are no such increasing or decreasing trends among all covariates, identifying a covariate as the special regressor becomes challenging. We do not explore this case here and intend to investigate it in future research.

For the identification of the merit parameters, the condition $\bm{1}^{\top}\bm{\theta}^*=0$ could also be used. However, our technique for theoretical analysis uses the inverse of
$\bm{V}=\bm{U}^\top \bm{U}$ in \eqref{def-V}, where $\bm{U}$ denotes the design matrix for $(\theta_1^*,\ldots,\theta_n^*)^\top$.
If $\theta_0^*$ is included, then $\bm{V}$ is not invertible.
Therefore, we use $\theta_0^*=0$ as the identification condition as in \cite{simons-yao1999}.
We established the consistency and asymptotic normality of the resulting estimators under some conditions.
Note that the conditions for guaranteeing asymptotic normality are stronger than those for consistency.
The asymptotic properties of the estimators depend not only on the width
but also on the eigenvalue of a projected design matrix.
It is of interest to see whether these conditions can be relaxed.

It is of interest to extend the methodology and theory to sparse comparison graphs such as  the Erd\"{o}s--R\'{e}nyi graph or heterogenous Erd\"{o}s--R\'{e}nyi graph \citep{han2024statistical, han2025unified}. Our proposed estimation method does not depend on the structures of comparison graphs and can be directly applied to any comparison graph. However, our techniques for theoretical analyses reply on properties of the complete comparison,
where the inverse of the information matrix $\bm{V}=\bm{U}^\top \bm{U}$ admits a closed form, with $\bm{U}$ denoting the design matrix of the merit parameters. In this case, $\bm{V}$ is a diagonally dominate matrix with positive diagonal elements and negative off-diagonal elements. In sparse comparison graphs, many entries of $\bm{V}$ are zero, which implies that $\bm{V}^{-1}$ does not have a closed form. One approach is to
approximate $\bm{V}^{-1}$ using a simple matrix \citep{simons-yao1999}.
\cite{Han-chen2020} gave the approximate error of the approximate inverse matrix under the Erd\"{o}s--R\'{e}nyi graph with edge probability $p$, where the error involves has a scalar factor $p^{-3}$.
Another approach is to apply the pseudoinverse technique developed by  \cite{han2025unified}. This requires a highly non-trivial analysis.  We would like to leave it for future research.
We further conducted simulations to evaluate the performance of the proposed estimator under sparse comparison graphs.
Following the setting of \cite{han2024statistical}, we simulate sparsity by modeling $n_{ij}\sim \text{Bernoulli}(T,p_{ij,n})$, where $p_{ij,n}$ is generated from the uniform distribution $U(p_n,q_n)$ with $p_n=1/\sqrt{n}$ and $q_n=p_n \log n $.
All other setting are the same as before.
The simulations results are in 
the Supplementary Material, which
indicate that the asymptotic properties of the estimators still hold in sparse paired comparison
settings.

\setlength{\itemsep}{-1.5pt}
\setlength{\bibsep}{0ex}
\bibliography{reference3}
\bibliographystyle{apa}

\end{document}